\documentclass[letterpaper, 10 pt, conference]{IEEEtran}

\usepackage{color}
\usepackage{multirow}
\usepackage{cite}
\usepackage{balance}

\usepackage{color}
\usepackage{multirow}
\usepackage{cite}
\usepackage{balance}
\usepackage{amssymb}

\usepackage{soul}

\usepackage{lipsum}
\usepackage[pdftex]{graphicx}
\usepackage[table,xcdraw]{xcolor}
\DeclareGraphicsExtensions{.pdf,.jpeg,.png,.JPG,}
\usepackage{graphicx}

\DeclareGraphicsExtensions{.eps}
\DeclareGraphicsExtensions{.tif}

\graphicspath{{figs/}}
\usepackage{epstopdf}
\usepackage{booktabs}

\usepackage{amsmath}
\usepackage{cleveref}
    \usepackage[caption=false, font=footnotesize]{subfig}

\begin{document}
	\title{\LARGE \bf Linear Model-Predictive Controller (LMPC) for Building's Heating Ventilation and Air Conditioning (HVAC) System}
	
	\author{Mohammad Ostadijafari, and Anamika Dubey
		\thanks{Authors are with School of Electrical Engineering and Computer Science,
        Washington State University, Pullman, WA 99164, USA}%
}
	\maketitle

	\begin{abstract}
		Model predictive control (MPC) is a widely used technique for temperature set-point tracking and energy optimization of Heating Ventilation and  Air  Conditioning (HVAC) systems in buildings. Unfortunately, a nonlinear thermal building model leads to a computationally expensive nonlinear MPC problem that is not suitable for real-time control and optimization. This paper presents a novel approximate linearized model for building's thermal dynamics and the HVAC system power consumption that leads to a computationally efficient linear model predictive controller (LMPC) for the buildings' HVAC systems. We employ feedback linearization technique to obtain an equivalent linearized model for the nonlinear thermal building dynamics and use constraint mapping approach to obtain a linearized formulation for new control variables. Next, using piecewise linearization, we obtain a linearized analytical model for the HVAC system power consumption. The proposed LMPC technique is validated using multiple simulation case studies. We demonstrate that the proposed LMPC technique is not only computationally efficient but also accurately approximates the nonlinear optimal control decisions.
	\end{abstract}

	\IEEEpeerreviewmaketitle
	\section{Introduction}
	\label{Introduction}
The building's primary  energy consumption is due to its Heating Ventilation and Air Conditioning (HVAC) system that is a controllable/flexible load. The embedded flexibility of operating HVAC system loads provides a multitude of opportunities to optimally control its operation for diverse requirements including but not limited to (1) minimization of the cost of electricity usage for the building, (2) maximization of customer's comfort level with a price/resource constraint, and (3) co-scheduling of HVAC systems with buildings' other distributed energy resources (DERs) and flexible loads. One of the problems that has been widely studied for the HVAC system controller design is a temperature set-point tacking problem where usually a Model-Predictive Controller (MPC) is used to track the desired temperature set-points by controlling the rate of the HVAC system airflow \cite{long2016hierarchical}. However, recently, due to opportunities for proactive demand-side participation, several researchers have explored the potential of MPC-based optimization approaches to meet an economic objective instead of set-point tracking for buildings' HVAC systems \cite{liu2018coordinating, parisio2013scenario}. 
	Note that for both optimal set-point tracking and economic optimization, the MPC problem essentially obtains an optimal temperature trajectory (for the building) based on the dynamical model for the HVAC system, occupants' desired comfort-level based on occupancy information or designated temperature set-points, forecasted values of external system variables including weather parameters, and other time-varying parameters, for example, price of electricity and/or profile for local DERs \cite{parisio2013scenario,safdarian2014comparison}. 
	
	
	The dynamical model required by a MPC controller must satisfy the following two requirements: the model should be (1) sufficiently accurate in describing system dynamics for a given set of variable parameters and (2) computationally tractable allowing for a real-time control and optimization \cite{cigler2013model,jafari2012frequency}. It is well known that building thermal dynamics and consequently the HVAC system model is highly nonlinear. To achieve a computational tractable MPC model, linearization of plant dynamics has drawn significant attention. For example, in \cite{maasoumy2012total,maasoumy2011model,wei2016proactive}, Jacobian linearization approach is used to eliminate the nonlinearity in the building thermal dynamical model. Authors then use the linear model to design a traditional MPC for temperature set-point tracking. In \cite{rehrl2011temperature}, authors use feedback linearization and develop MPC technique to track a set-point temperature using water-to-air heat exchange in HVAC systems. Unfortunately, these methods are only valid when room temperature is allowed to vary within a very small interval that is usually the case for set-point tracking problems. However, economic optimization may require larger variations in room temperature especially when building's occupancy information can be leveraged to improve efficiency. Note that when the building is not occupied at certain time of the day, it can be overheated or overcooled to achieve the desired economic objective. To address this problem, two methods have been explored in related literature: (1) solving a full non-linear MPC \cite{ma2012model}; (2) a mixed-integer approximation of non-linear problem by discretizing the states of control variable i.e. air mass flow rate \cite{liu2018coordinating}. For example, authors in \cite{liu2018coordinating} formulate a MILP model based on the nonlinear plant dynamics to minimize the cost of electricity usage by the HVAC system while satisfying the occupants' comfort level. 
	
	
Owing to model non-linearities, to optimally achieve a desired objective, the HVAC system requires a nonlinear model-predictive controller (NLMPC). Unfortunately, a NLMPC is computationally expensive and therefore, the approach is not suitable for real-time optimization and control. The need for computational tractability calls for novel linearization techniques that can closely mimic the nonlinear plant dynamics. The objective of this paper is to develop novel linearization techniques that unlike Jacobian-linearization, can accurately mimic the system nonlinearities without approximating the trajectory for any decision-variable. We use a combination of linearization techniques including feedback linearization, constraint mapping, and piecewise linearization to simultaneously approximate the nonlinearity in a building thermal dynamical model and the HVAC system power consumption equations. This leads to a linearized model predictive controller (LMPC). Using simulations, we validate that the proposed approach results in a controller of reduced complexity that accurately approximates the original nonlinear plant dynamics and economic constraints. 

	
	\vspace{-0.2cm}
	\section{Dynamical Model for Building Thermal Load}
	\vspace{-0.1cm}
	\label{System model}
	Thermal dynamical model of a building is usually obtained by modeling the building as a first-order RC network  \cite{maasoumy2011model,maasoumy2012total}. In the resulting RC network, a node indicates a wall or a room. In general, if there are in total $n$ nodes, $m$ of which denote rooms, then, $n-m$ remaining nodes denote walls \cite{maasoumy2014handling}. 
	In order to design a practical controller, we discretize governing equations of rooms and walls using zero-hold discretization \cite{liu2018coordinating}. The temperature of $i^{th}$ wall is characterized in (\ref{eqa})
	\vspace{-0.2cm}
	\begin{equation}\label{eqa}
	\small
	c_{w_i}(T^{k+1}_{w_i}-T^{k}_{w_i})=\tau\left[\sum_{j\in adj(w_i)}{\frac{T^{k}_{j}-T^{k}_{w_i}}{R^{'}_{ij}}+r_i \alpha_i A_i q^{k}_{rad_i}}\right]
	\end{equation}
	where $c_{w_i}$, $\alpha_i$ and $A_i$ are the heat capacity, absorption coefficient and area of wall $i$, respectively; $\tau$ is the length of time interval between two consecutive sampling time; $T^{k}_{w_i}$ and $q^{k}_{rad_i}$ are the surface temperature and the solar radiation density of wall $i$ at the sampling time $k$, respectively; $adj(w_i)$ is the set of nodes which are adjacent to node $w_i$; $T^{k}_{j}$ is the temperature of the $j^{th}$ adjacent node at the sampling time $k$; $R^{'}_{ij}$ is the total resistance between wall $i$ and node $j \in adj(w_i)$; and $r_i=0$ for internal walls, $r_i=1$ for peripheral walls. The temperature of $i^{th}$ room is governed by (\ref{eqb}).
	\begin{equation}\label{eqb}
	\small\vspace{-0.2cm}
	\begin{split}
    c_{r_i}(T^{k+1}_{r_i}-T^{k}_{r_i})& ={\tau}[\sum_{j\in adj(r_i)}{\frac{T^{k}_{j}-T^{k}_{r_i}}{R^{'}_{ij}}+u^{k}_{i}c_a(T_{s_i}-T^{k}_{r_i})}\\& +g_i{\beta}_{win_i} A_{win_i}q^{k}_{rad_i}+q^{k}_{int_i}]
	\end{split}
	\vspace{-0.1cm}
	\end{equation}
	where $c_{r_i}$ and $T_{s_i}$ are the heat capacity and the temperature of supply air of room $i$, respectively; $T^{k}_{r_i}$ are $u^{k}_{i}$ are the indoor temperature and air mass flow to the room $i$ at the sampling time $k$, respectively; $adj(r_i)$ is the set of nodes which are adjacent to node $r_i$ and have connection with it (e.g. walls inside of the room $i$ or rooms that are separated by a wall and affect each other); $c_a$ is the specific heat capacity of the air; $\beta_{win_i}$ is the transmissivity of glass of the window in room $i$; $ A_{win_i}$ is the total area of window on surrounding walls in room $i$; $q^{k}_{rad_i}$ is the solar radiation density radiated from the window to the room $i$ at sampling time $k$; and $g_i=0$ if there is no window in room $i$ otherwise $g_i=1$.
	\vspace{-0.2cm}
	\subsection{State Space formalization of the Building Thermal Model}
	After deriving the above heat transfer equations for all $n$ nodes in the equivalent RC network of the building, we obtain the following state-space equations representing the building thermal model:
	\vspace{-0.2cm}
	\begin{equation}\label{eq1}
	\small
	\boldsymbol{x}^{k+1}=\boldsymbol{A}\boldsymbol{x}^{k}+\boldsymbol{B}\boldsymbol{u}^{k}\circ(\boldsymbol{T_s}-\boldsymbol{y}^{k})+\boldsymbol{E}\boldsymbol{d}^{k}
	\end{equation}
	\begin{equation}\label{eq2}
	\small
    \boldsymbol{y}^{k}=\boldsymbol{C}\boldsymbol{x}^{k}
	\end{equation} 
	where superscript $k$ shows the sampling time and $\circ$ is the element-wise product operator for two vectors; $\boldsymbol{d}^{k} \in \mathbb{R}^{l}$ is the vector of disturbance (with $l$ number of the disturbance elements such as external temperature, solar radiation and internal gains, etc.) at sampling time $k$; $\boldsymbol{A} \in \mathbb{R}^{n\times n}$, $\boldsymbol{B} \in \mathbb{R}^{n\times m}$, $\boldsymbol{C} \in \mathbb{R}^{m\times n}$ and $\boldsymbol{E} \in \mathbb{R}^{n\times l}$ are matrices obtained from a building thermal model representing time-invariant building parameters (see \cite{maasoumy2012total,haghighi2011modeling} for more details); $\boldsymbol{x}^{k} \in \mathbb{R}^{n}$ is the state vector representing the temperature of the network nodes; $\boldsymbol{u}^k \in \mathbb{R}^{m}$ is the vector of input variables whose elements ($u^{k}_i$) are air mass flow into each thermal zone; $\boldsymbol{y}^{k} \in \mathbb{R}^{m}$ is the output vector of the system; and $\boldsymbol{T_s}\in \mathbb{R}^{m}$ with entries of $T_{s_i}$ representing the temperature of supply air for different rooms of the building.
\subsection{HVAC System Power Consumption Model}
	Next, we detail the equations for power consumption of the HVAC system as a function of air mass flow rate. A typical HVAC system consumes most of its power through the heater, chiller, and fan \cite{liu2018coordinating}. Without loss of generality, in this paper we only consider a cooling system. The fan power consumption, $P^{k}_{f_i}$, is modeled as a cubic function of air mass flow rate, $u_i^k$.
	\begin{equation}\label{eq3}
	P^{k}_{f_i}=P_{{rated}_i} (u^{k}_i/u_{{rated}_i})^3
	\end{equation}
	where $P_{{rated}_i}$ and $u_{{rated}_i}$ are the rated power and rated outlet air mass flow rate of the air handling unit of a HVAC system in thermal zone $i$, respectively; and $P_{f_i}^k$, $u^{k}_i$ are power consumption and the air mass flow rate (control variable) of fans in thermal zone $i$ at sampling time $k$, respectively. 
	
	The cooling load is a function of the air mass flow rate, ambient temperature, and temperature of the thermal zone $i$  as defined in \cite{liu2018coordinating}:
		\vspace{-0.2cm}
	\begin{equation}\label{eq4}
	\small
	P^{k}_c=\frac{c_a}{COP}\sum^{m}_{i=1}{u^{k}_i\left[d_p y^k_i+(1-d_p)T^k_{out}-T_{s_i}\right]} 
	\vspace{-0.2cm}
	\end{equation}
	where $T^k_{out}$ is the ambient temperature at sampling time $k$; $COP$ is a performance coefficient of the chiller; $c_a$ is the specific heat capacity of the air; and $d_p$ is the instantaneous return-to-total ratio of the chiller that varies between $0$ and $1$. Therefore, the total power consumption of the entire building by its the HVAC system at sampling time $k$ is given by (\ref{eq5}).
	
	\vspace{-0.3cm}
	\begin{equation}\label{eq5}
	\small
	 P^{k}_H=P^{k}_c+\sum^{m}_{i=1}P^{k}_{f_i}
	 \vspace{-0.2cm}
	\end{equation}
	\subsection{System nonlinearity}
	\label{System nonlinearity}
	The objective of this paper is to propose a generalized MPC controller for a HVAC system that is computationally tractable while can be used to meet diverse system-level objectives. MPC is a model-based controller that requires the dynamical model of the system to obtain optimal control inputs. The required model of the system must be sufficiently accurate to acquire a valid prediction of system states in a computationally tractable manner \cite{cigler2013model}. The building thermal model dynamics illustrated by (\ref{eq1})-(\ref{eq5}) is nonlinear. In this section we describe several nonlinearities in the model and why they cannot be approximated using traditional Jacobian-based linearization approach when optimizing for a general objective function.
	
	\begin{itemize}
	\item {\em{Building Thermal Dynamical Model} (Bilinear term in constraint (\ref{eq1}))}: The main nonlinearity in the system model is due to a bilinear term -- the product of system inputs and output variables ($u^k_i. y^k_i$) in  (\ref{eq4}). Note that desirable system outputs are rooms' temperature. ($\boldsymbol{y}^{k}$=$\boldsymbol{T}^{k}_{r}$) which are also system state variables. Therefore, the dynamical system model is  bilinear in the system's control inputs and  output variables which cannot be linearized using Jacobian-linearization techniques especially, when occupancy information is leveraged to meet an efficiency/economic objective. In this case the room temperature can vary over a wide-range (to overheat/overcool the building) rendering jacobian-linearization inapplicable.  
	\item {\em{Fan Power (Cubic control input term in (\ref{eq3}))}}: The second nonlinearity originates from the equation for fan power that is a cubic function of control variable i.e. air mass flow rate, $(u^{k}_i)$. This nonlinearity is critical when optimizing for power consumption of the HVAC system subject to dynamical model and occupancy information and hence, unlike temperature set-point tracking problem cannot be ignored. \item {\em{Chiller Power (Bilinear term in  (\ref{eq4}))}}: The expression for chiller power described in  (\ref{eq4}) is also nonlinear due to bilinear product of control input and output variables, $u^k_i. y^k_i$. This nonlinearity is critical when optimizing for power consumption of the HVAC system provided that temperature set-points and optimal trajectory for room temperature can vary over a wide-range. 
\end{itemize}
\section{Model Predictive Control of the HVAC System}
A nonlinear MPC control (based on the nonlinear building thermal model dynamics) is computational intensive and may not be practical for real-time control, especially when co-scheduling a large-number of buildings' flexible resources with time-varying price signals. Note that temperature set-points can vary over a wide-range due to varying occupancy patterns when attempting to optimize electricity usage for time-varying cost of electricity or/and maximizing the building occupants' comfort level. Under these scenarios, the building is not occupied at certain time of the day and/or the desirable temperature set-point of occupants vary significantly during different time intervals. Thus, to ensure an optimal use of electricity for the HVAC system, a nonlinear MPC problem needs to be solved that can vary the temperature set-point trajectory over a wide range. In this section we describe two distinct optimization problems that leverage occupancy information to optimize the operation of building's HVAC system. 


\subsection{Minimize Transacted Energy Cost for the HVAC System}
The problem objective is to optimally control the HVAC system such that it can optimize the net cost of transacted energy for the specified prediction window while ensuring that the desired level of comfort is met for its consumers. We formulate the problem as a model predictive control (MPC) problem with the objective of minimizing the building's total electricity usage cost for a given price vector as indicated in time-of-use (TOU) electricity tariffs for each hour of the day subject to thermal dynamical models and load satisfaction constraints.
	
	\vspace{-0.3cm}
	\begin{small}
    \begin{equation}\label{eq6}
	\underset{\boldsymbol{u}^{k}}{Min}\sum^{t+W-1}_{k=t}{{Price}^{k}.P^{k}_{H}}
	\end{equation}
	\end{small}
	Subject to:
    \vspace{-0.1cm}
	\begin{small}
    \begin{equation}\label{eq7}
	T^{k}_{Min}\leqslant T^{k}_r \leqslant T^{k}_{Max}
	\end{equation}
	\end{small}
	    \vspace{-0.3cm}
	\begin{small}
    \begin{equation}\label{eq8}
	0\leqslant P^{k}_H \leqslant P^k_{H_{Max}}
	\end{equation}
	\end{small}
    \vspace{-0.3cm}
	\begin{small}
    \begin{equation}\label{eq9}
	u^{k}_{Min}\leqslant u^{k} \leqslant u^{k}_{Max}
	\end{equation}
	\vspace{-0.5cm}
	\begin{align*}
	\text{constraints (\ref{eq1}) - (\ref{eq5})}   
	\end{align*}
	\end{small}
	The minimization of the electricity usage cost is given by (\ref{eq6}), where $W$ is the prediction window, and $Price^k$ is the electricity tariff at the sampling time $k$. The desired temperature range, the HVAC system power consumption limits, air mass flow limits, the thermal building model, and the total power consumption by the HVAC system are presented in constraints (\ref{eq7}), (\ref{eq8}), (\ref{eq9}), [(\ref{eq1}) and (\ref{eq2})], and [(\ref{eq3})-(\ref{eq5})], respectively;  where at sampling time $k$, $T^{k}_{Min}, T^{k}_{Max}, P^{k}_{H_{Max}}, u^{k}_{Min}$ and $u^{k}_{Max}$ represent minimum and maximum range for building temperature ($^{\circ}C$), maximum the HVAC system power consumption limit, and minimum and maximum limits for the HVAC system air mass flow rate, respectively. 
	
	
\vspace{-0.2cm}	
\subsection{Budget-Constrained Occupants' Comfort Maximization}
\vspace{-0.1cm}
The problem objective is to maximize occupants' comfort by maintaining the building temperature close to their desired temperature trajectory, $T_{oc}^k$, while including the building occupancy information. The optimization problem is constrained by the total available budget for the HVAC system energy usage, $B$. Occupants provide a desired trajectory for building temperature based on their preferred temperature level and occupancy information. The problem formulation is described as follows: 

	\vspace{-0.1cm}
	\begin{small}
    \begin{equation}\label{eq12}
	\underset{\boldsymbol{u}^{k}}{Max}\sum^{t+W-1}_{k=t}{|T_r^k - T^k_{oc}|}
	\end{equation}
	\end{small}
	Subject to:
	\begin{small}
	\begin{equation}\label{eq16-a}
	\sum^{t+W-1}_{k=t}{{Price}^{k}.P^{k}_{H}} \leq B
	\end{equation}
	\end{small}
	\vspace{-0.5cm}
	\begin{align*}
	\text{constraints (\ref{eq1}) - (\ref{eq5}), (\ref{eq8}) - (\ref{eq9})}   
	\end{align*}
Here, (12) represents the problem objective of maximizing occupants' comfort and (13) describes the budget constraint. Similar to the problem description in III.A, the control problem is constrained by building thermal dynamical model in (3)-(7), and constraints on system variables in (10)-(11).

The nonlinearity in constraints for both problems requires a NLMPC to solve the above control problem. Note that, Jacobian-linearization is not valid for this problem as  temperature set-points can vary over a wide-range due to varying occupancy patterns when attempting to optimize HVAC system's electricity consumption. In this paper, we propose a novel Linearized-MPC (LMPC) model that is not only computationally efficient with improved processing time, but also providesa valid approximation for the original nonlinear plant dynamics for large variations in room-temperature due to occupancy changes. Specifically, we employ the following three techniques to accurately approximate the plant model: (1) feedback-linearization for bilinear constraint in the building thermal dynamical model \cite{khalil2002nonlinear}, (2) constraints mapping to new decision variables space, and (3) piecewise linearization to linearize cubic power relationship. Since we do not make any assumption regarding desired temperature trajectory, the proposed linearized formulation is able to optimally control building's HVAC system in a computationally tractable manner for a wide-range of economic objectives. 

\section{Linearized Model Predictive Controller}
\label{LMPC}
The proposed linearized model for optimal control of buildings’ HVAC system is detailed in this section. 
First, an exact linear model for the dynamical HVAC system model is obtained using a feedback linearization technique. Next, the nonlinear relationship between power consumption of fan and chiller and air mass flow rate is linearized using piecewise linearization technique. This results in a linear MPC formulation that can be solved using off-the-shelf linear optimization software packages. The resulting model is not only computationally efficient but also closely approximates the nonlinear plant dynamics of building's HVAC system.

	
	\subsection{Feedback Linearization of Bilinear HVAC Dynamics}
\label{FLB}
 To handle non-linearity in (\ref{eq1}), feedback linearization is proposed in this section where by finding an explicit relation between the system outputs $\boldsymbol{y}^{k}$ and the control inputs $\boldsymbol{u}^{k}$, we cancel out the non-linearity of the system. Interested readers can refer to \cite{khalil2002nonlinear} and \cite{koorehdavoudi2018impacts} for further details regarding feedback linearization and feedback control approaches. Therefore, to linearize (\ref{eq1}), we define the following equation:
    \vspace{-0.1 cm}
    \begin{equation}\label{eq10}
	\small
	\boldsymbol{v}^{k}=\boldsymbol{u}^{k}\circ(\boldsymbol{T_s}-\boldsymbol{y}^{k})
	\vspace{-0.1 cm}
	\end{equation}
	where $\boldsymbol{v}^k \in \mathbb{R}^{m}$ is the vector of new input variables at sampling time $k$ after feedback linearization whose elements ($v^{k}_{i}$) are new input variables for thermal zone $i$. 
	
	Using (\ref{eq10}) in (\ref{eq1}), we obtain:
	\begin{equation}\label{eq11}
	\small
	\boldsymbol{x}^{k+1}=\boldsymbol{A}\boldsymbol{x}^{k}+\boldsymbol{B}\boldsymbol{v}^{k}+\boldsymbol{E}\boldsymbol{d}^{k}
	\vspace{-0.1 cm}
	\end{equation}
	It should be noted that in (\ref{eq11}), state vector $\boldsymbol{x}^{k}$ remains the same as the one in (\ref{eq1}), and as a result the same equation (\ref{eq2}) also describes the output equation for the building thermal model after implementing the feedback linearization approach. 
	
	Note that since (\ref{eq11}) is based on the new control input vector, the minimization problem defined in (\ref{eq6}) should change accordingly. The following equation defines the minimization of electricity usage cost obtained after feedback linearization:
	\begin{equation}\label{eq12-a}
	\small
	\underset{\boldsymbol{v}^{k}}{Min}\sum^{t+W-1}_{k=t}{{Price}^{k}.P^{k}_{H}}
	\end{equation}
	
	The next step is to map the constraints (\ref{eq3})-(\ref{eq5}) and (\ref{eq9}), which are based on $\boldsymbol{u}^{k}$ to the new input variables $\boldsymbol{v}^{k}$ as detailed in the following section. 
\vspace{0.1cm}	
\subsection{Mapping Input Constraints}
\label{MIC}
In this section, we address the mapping of constraint (\ref{eq9}) to the new input variable $\boldsymbol{v}^{k}$. The mapping problem, at each sampling time $k$, can be defined as deriving the constraints of the following form: 
    \begin{equation}\begin{aligned}\label{eq13}
    \small
	\boldsymbol{v}^{k+j|k}_{Min}\leqslant \boldsymbol{v}^{k+j|k}\leqslant  \boldsymbol{v}^{k+j|k}_{Max} \hspace{0.2cm} \text{for} \hspace{0.1cm} j= 0, 1,..., W-1 
	\end{aligned}
	\end{equation}
where $\boldsymbol{v}^{k+j|k}$ is the value of LMPC control input vector $\boldsymbol{v}^{k+j}$ (for future time $k+j$) obtained at sampling time $k$, and $\boldsymbol{v}^{k+j|k}_{Min}$ and $\boldsymbol{v}^{k+j|k}_{Max}$ are the minimum and maximum limit for control input $\boldsymbol{v}^{k+j}$ obtained at sampling time $k$, respectively. 

This transformation process should be solved at each sampling time, as the mapping is output dependent. That is, for finding the constraints in (\ref{eq13}) at each sampling time, mapping is performed by solving an optimization problem defined in (\ref{eq14}) and output vector at current sampling time, $\boldsymbol{y}^k$. The resulting optimization problem is defined as the following:
\begin{equation}\begin{aligned}\label{eq14}
	\small
\boldsymbol{v}^{k+j|k}_{Min}=\underset{\boldsymbol{u}^{k+j|k}}{Min}   [\boldsymbol{u}^{k+j|k}\circ(\boldsymbol{T_s}-\boldsymbol{y}^{k+j|k})] \\
\boldsymbol{v}^{k+j|k}_{Max}=\underset{\boldsymbol{u}^{k+j|k}}{Max}  [\boldsymbol{u}^{k+j|k}\circ(\boldsymbol{T_s}-\boldsymbol{y}^{k+j|k})] 
	\end{aligned}\end{equation}
\vspace{-0.4cm}	\noindent subject to 
	\begin{equation}\begin{aligned}
\vspace{-0.4cm}	\small
\nonumber	\boldsymbol{u}^{k}_{Min}\leqslant \boldsymbol{u}^{k+j|k} \leqslant \boldsymbol{u}^{k}_{Max} \hspace{0.15cm} \text{for} \hspace{0.1cm} j= 0, 1,..., W-1 
	\end{aligned}\end{equation}
	where $\boldsymbol{u}^{k+j|k}$ and $\boldsymbol{y}^{k+j|k}$ represent the input $\boldsymbol{u}^{k+j}$ and output $\boldsymbol{y}^{k+j}$ computed at sampling time $k$. By substituting $j=0$ in (\ref{eq14}), the constraints on the input control at the current sampling time can be calculated as:
	\begin{equation}\begin{aligned}\label{eq15}
	\small
	\boldsymbol{v}^{k|k}_{Min}=\underset{\boldsymbol{u}^{k}}{Min} \: \:  \boldsymbol{u}^{k}\circ(\boldsymbol{T_s}-\boldsymbol{y}^{k})\\
	\boldsymbol{v}^{k|k}_{Max}=\underset{\boldsymbol{u}^{k}}{Max}  \: \: \boldsymbol{u}^{k}\circ(\boldsymbol{T_s}-\boldsymbol{y}^{k})
	\end{aligned}\end{equation}
	\noindent subject to 
	 \begin{equation}
	    \small
\nonumber	\boldsymbol{u}^{k}_{Min}\leqslant \boldsymbol{u}^{k} \leqslant \boldsymbol{u}^{k}_{Max}
	\end{equation}

	As is clear, this optimization problem is trivial to solve due to affine objective function in $u^k_i$ \cite{henson1997feedback}. On the other hand, it is difficult to compute the constraints for future inputs over the prediction window $[\boldsymbol{v}^{k+1|k}, \boldsymbol{v}^{k+2|k},..., \boldsymbol{v}^{k+W-1|k}]$. Note that in order to solve the optimization problem formulated in (\ref{eq14}) to obtain the mapped constraints in (\ref{eq13}), the estimates of the future values of input and output variables are needed. However, these estimates are not available until the LMPC problem is solved that in turn requires the mapped input constraints in (\ref{eq13}) over the entire prediction window \cite{kurtz1997input}. To address this problem, we use a similar but slightly modified approach illustrated in \cite{ghasemi2011application} described as follows. 

At the first sampling time of solving the problem ($k=0$), we use the constant input constraints to calculate the bounds on future control inputs in prediction window $W$ as follows:
    \begin{equation}\begin{aligned}\label{eq16}
	\small
	\boldsymbol{v}^{k+j|k}_{Min}=\boldsymbol{v}^{k|k}_{Min} \\
	\boldsymbol{v}^{k+j|k}_{Max}=\boldsymbol{v}^{k|k}_{Max}
	\end{aligned}
	\end{equation}
	\vspace{-0.25cm}
	\begin{align*}
	\small
	\text{for}\: \:  j= 1, 2,..., W-1  
	\end{align*}
	where $\boldsymbol{v}^{k|k}_{Min}$ and $\boldsymbol{v}^{k|k}_{Max}$ are obtained based on (\ref{eq15}). Hence, LMPC uses (\ref{eq15}) and (\ref{eq16}) to solve the optimization problem formulated in (\ref{eq12}) at  the first sampling time ($k=0$). Then, for solving the problem at each of the next sampling times $(k=1, 2, ..., W-1)$, we use inputs calculated from the previous sampling time to calculate the future constraints at the current sampling time. The resulting problem is formulated as following:
	 \begin{equation}\begin{aligned}\label{eq17}
	 \small
	\boldsymbol{v}^{k+j|k}_{Min}=\underset{\boldsymbol{u}^{k+j|k}}{Min}   [\boldsymbol{u}^{k+j|k-1}\circ(\boldsymbol{T_s}-\boldsymbol{y}^{k+j|k-1})] \\
	\boldsymbol{v}^{k+j|k}_{Max}=\underset{\boldsymbol{u}^{k+j|k}}{Max}  [\boldsymbol{u}^{k+j|k-1}\circ(\boldsymbol{T_s}-\boldsymbol{y}^{k+j|k-1})] \\
	\end{aligned}
	\end{equation}
	\vspace{-0.4cm}
	\begin{align*}
	\small
	\text{for}\: \:  j= 1, 2,..., W-1  
	\end{align*}
	\vspace{-0.2cm}
	\noindent subject to: 
	\begin{equation*}
	\small
	\boldsymbol{u}^{k}_{Min}\leqslant \boldsymbol{u}^{k+j|k} \leqslant \boldsymbol{u}^{k}_{Max}\hspace{0.2cm} \text{for} \hspace{0.1cm} j= 0, 1,..., W-1 
	\end{equation*}
	
In (\ref{eq17}), for each thermal zone $i$, if $({T_{s_i}}-{y}^{k+j|k-1}_i)> 0$, then ${u}^{k}_{Min,i}$ and ${u}^{k}_{Max,i}$ determines $\boldsymbol{v}^{k+j|k}_{Min,i}$ and ${v}^{k+j|k}_{Max,i}$, respectively; otherwise  ${u}^{k}_{Min,i}$ and ${u}^{k}_{Max,i}$ determines ${v}^{k+j|k}_{Max,i}$ and ${v}^{k+j|k}_{Min,i}$, respectively. Note that the bounds for the control inputs at the current sampling time are obtained using (\ref{eq15}).
\vspace{-0.2cm}
\subsection{Linearized Power Consumption Model for HVAC}
\vspace{-0.1cm}
As previously mentioned, constraints (\ref{eq3}), (\ref{eq4}) and consequently (\ref{eq5}) introduced in Section II are based on $\boldsymbol{u}^k$. After using feedback linearization technique, these should be redefined based on the LMPC control input, $\boldsymbol{v}^k$. 
	
First, we consider constraint (\ref{eq3}). By modifying constraint (\ref{eq10}) at each sampling time $k$, we obtain $\boldsymbol{u}^k$ based on $\boldsymbol{v}^k$ over the prediction window in (\ref{eq18}).
    \begin{equation}\begin{aligned}\label{eq18}
    \small
    \boldsymbol{u}^{k+j|k}=\frac{\boldsymbol{v}^{k+j|k}}{\boldsymbol{T_s}-\boldsymbol{y}^{k+j|k}} \hspace{0.1cm} \text{for} \hspace{0.1cm} j= 0, 1,..., W-1 
	\end{aligned}
	\end{equation}
	\noindent where for each sampling time $k$ and for each $j$, each entry of the right hand-side of the above equation is defined as $\frac{{v}^{k+j|k}_i}{{T_{s_i}}-{y}^{k|k+j}_i}$. 
	Substituting (\ref{eq18}) in (\ref{eq3}), we obtain the following:
	\begin{equation}\label{eq19}
	\small
	{P}^{k+j|k}_{f_i}=P_{{rated}_i} \left(\frac{{v}^{k+j|k}_i}{{(T_{s_i}}-{y}^{k+j|k}_{i}) u_{{rated}_i}}\right)^3 \text{for } j= 0, ... W-1  
	\end{equation}
	where ${P}^{k+j|k}_{f_i}$, is the total fan power consumption ${P}^{k+j}_{f}$ computed at sampling time $k$ for the thermal zone $i$. Although (\ref{eq19}) is based on the input variable of LMPC, it is non-linear due to cubic relation between ${P}^{k+j|k}_{f_i}$ and ${v}^{k+j|k}_i$, and inverse-polynomial relation between ${P}^{k+j|k}_{f_i}$ and ${y}^{k+j|k}_i$; hence, it cannot be directly integrated into the LMPC model. To eliminate the nonlinearity between ${P}^{k+j|k}_{f_i}$ and ${y}^{k+j|k}_i$, we use the same method we previously proposed in Section \ref{MIC} for mapping input constraints as detailed below. 
	
	For the first sampling time ($k=0$), we consider the initial value as $\boldsymbol{y}^{0} \in \mathbb{R}^{m}$ (with entries of $y^0_i$ for the vector $\boldsymbol{y}^{k}$), and consider it as the value of the output for the rest of the sampling time in the prediction window as follows:
	    \begin{equation}\begin{aligned}\label{eq20}
	\small
	\boldsymbol{y}^{k+j|k}=\boldsymbol{y}^{0} \hspace{0.1cm} \text{for} \hspace{0.1cm} j= 0, 1,..., W-1 
	\end{aligned}
	\end{equation}

	
	Then, for the next sampling times in the prediction window, i.e. $(k=1, 2, ..., W-1)$, we use the output calculated at the previous sampling time, $\boldsymbol{y}^{k+j|k-1}$ as the future outputs at the current sampling time, $\boldsymbol{y}^{k+j|k}$, stated as the following:
	    \begin{equation}\begin{aligned}\label{eq21}
	\small
	\boldsymbol{y}^{k+j|k}=\boldsymbol{y}^{k+j|k-1} \hspace{0.1cm} \text{for} \hspace{0.1cm} j= 0, 1,..., W-1 
	\end{aligned}
	\end{equation}

Note that using this approach, future output variables ${y}^{k+j|k}_i$ are constant and equal to the output variables obtained at the previous sampling time.  This approach, therefore, makes ${y}^{k+j|k}_i$ constant at sampling time $k$ and eliminates the inverse-polynomial relation between ${P}^{k+j|k}_{f_i}$ and ${y}^{k+j|k}_i$. Although approximate, this method successively improves the prediction of output variables at the current sampling time.

Note that (\ref{eq4}) is also nonlinear due to the product of variables in form of $u^k_i. y^k_i$.   This imposes the term $v^k_i. y^k_i$ in the equivalent chiller  equation after feedback linearization. This nonlinearity is also relaxed using the same approach proposed at the beginning of this section by modifying $\boldsymbol{y}^{k+j|k}$ based on (\ref{eq20}) and (\ref{eq21}), detailed as follows:

\vspace{-0.4cm}
	\begin{small}
	\begin{equation*}\begin{aligned}\label{eq22}
	P^{k}_c=\frac{c_a}{COP}\sum^{m}_{i=1}{\left(\frac{{v}^{k}_{i}}{{T_{s_i}}-{y}^{0}}\right)\left(d_p y^k_i+(1-d_p)T^k_{out}-T_{s_i}\right)} 
	\end{aligned}\end{equation*}
		\vspace{-0.3cm}
	\begin{align}\label{eq34}
	\text{for}\: \:  j= 0  
	\end{align}
	\vspace{-0.6cm}
	\begin{multline*}
	\small
	 P^{k+j|k}_c=\frac{c_a}{COP}\sum^{m}_{i=1}{\left(\frac{v^{k+j|k}_{i}}{T_{s_i}-y^{k+j|k-1}_i}\right)}\\   .\left(d_p{y}^{k+j|k-1}_i+(1-d_p)T^k_{out}-T_{s_i}\right)
	\end{multline*}
		\vspace{-0.5cm}
	\begin{align*}
	\nonumber \text{for}\: \:  j= 1, 2,..., W-1  
	\end{align*}
	\end{small}

\vspace{-0.6cm}
Next, we employ the incremental approach of piecewise linearization to relax the nonlinearity in $\boldsymbol{P}^{k+j|k}_{f}$ due to cubic term of $\boldsymbol{v}^{k|k+j}$. The approach is explained briefly here; however, interested readers can refer to \cite{bernreuther2017solving,bussieck2003mixed} for further details on the piecewise linearization approach.

At each sampling time, we approximate each ${v}^{k+j|k}_i$ in (\ref{eq13}) as the summation of multiple line segments. In order to formulate the linear approximated function, we first introduce following two conditions \cite{bernreuther2017solving}:
	\begin{itemize}
	    \item The line segments in  $\mathbf{L}$ can be ordered as $l_1, l_2, ...,l_{\mathbf{L}}$ subject to $l_j \cap l_{j-1} \ne$\o \ for $j \in \{2,...,|\mathbf{L}|\}$, where operator $|.|$ shows the number of the elements.
	    \item For the order in $\mathbf{L}$, the vertices of each line segment $l_j$  ordered as $h^0_j$ and $h^1_j$ for $j \in \{2,...,|\mathbf{L}|\}$. 
	\end{itemize}
	
     At each sampling time, the limits of the linearized control input specified by (\ref{eq19}) determines the initial vertex  of the first line segment ($h^0_1$) and ending vertex of the last line segment ($h^{1}_\mathbf{L}$).  Then, by introducing auxiliary continuous variables $\sigma_j$, which can be interpreted as the slopes of line segments $l_j$ with the vertices $h^0_j$ and $h^1_j$ for $j \in \{1,...,|\mathbf{L}|\}$, we can define the piecewise linearized input $v^k_i$ as follows:

\vspace{-0.3cm}
	\begin{small}
	\begin{eqnarray}\label{eq23}
\nonumber	v^k_{i}=h^0_1+\sum^{|\mathbf{L}|}_{j=1}{\sigma_j(h^1_j-h^0_j)}\\
	f(v^k_i)=f(h^0_1)+\sum^{|\mathbf{L}|}_{j=1}{\sigma_j[f(h^1_j)-f(h^0_j)]}\\ 
\nonumber	 \sigma_j\geqslant 0  \hspace{0.1cm}\text{for} \: j \in \{1,...,|\mathbf{L}|\}
	\end{eqnarray}
	\end{small}
	\vspace{-0.4cm}
	
	 \noindent where $f(.)=(.)^3$ for any arbitrary value of $(.)$ in (\ref{eq23}). Note that, as ${(v^k_i)}^3$ is monotonically increasing function,  there is no need for binary variables to define piecewise linear model. 
	 Using (\ref{eq20}), (\ref{eq21}) and (\ref{eq23}), we linearize (\ref{eq19}) as following:
	 
	 \vspace{-0.3cm}
	 \begin{small}
	 \begin{equation}\begin{aligned}\label{eq24}
	 {P}^{k}_{f_i}&=&P_{{rated}_i} \left(\frac{1}{({T_{s_i}}-{y}^{0}_i) u_{{rated}_i}}\right)^{3}f_(v^k_i) \hspace{0.1cm} \text{for} \hspace{0.1cm} j=0\\ {P}^{k+j|k}_{f_i}&=&P_{{rated}_i}\left(\frac{1}{({T_{s_i}}-{y}^{k+j|k-1}_i) u_{{rated}_i}}\right)^{3}f(v^{k+j|k}_i)\\
	\end{aligned}\end{equation}
		\vspace{-0.5cm}
	\begin{align*}
	\text{for}\: \:  j= 1, 2,..., W-1  
	\end{align*}
	\end{small}
	\vspace{-0.3cm}

 
\vspace{-0.3cm}

\begin{figure}[h]
		\centering
			\vspace{-0.4cm}
		\includegraphics[width=1\linewidth]{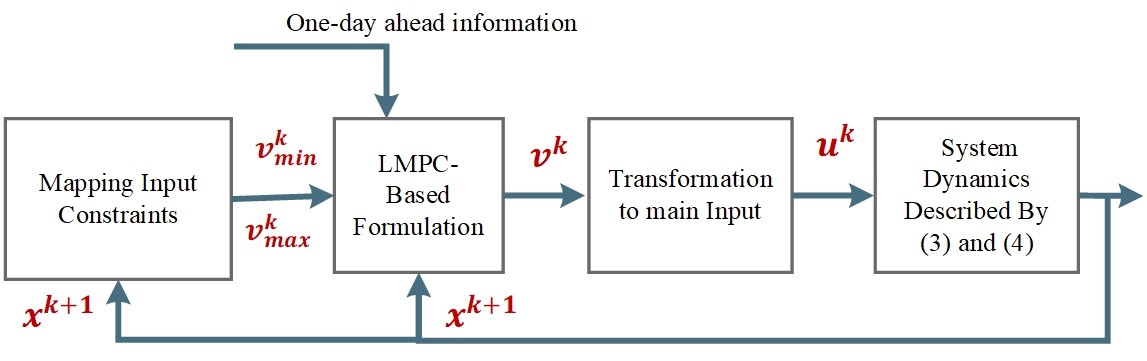}
		\vspace{-0.6cm}
		\caption{Schematic for the Proposed LMPC controller}
		\label{fig:2}
		\vspace{-0.4cm}
	\end{figure}

The LMPC problem is illustrated in Fig. \ref{fig:2}. At the beginning of each day,  LMPC controller receives one-day ahead prediction information. Then, at each sampling time $k$, using this information and mapped input constraints defined in (\ref{eq14}), the LMPC solves the optimization problems detailed in Section III with all constraints converter to linear form using proposed methods in Section IV. This results in a vector of feedback linearized input, $[\boldsymbol{v}^{k}, \boldsymbol{v}^{k+1}, \boldsymbol{v}^{k+2},..., \boldsymbol{v}^{k+W-1}]$. The first entry of linearized control input trajectory ($\boldsymbol{v}^{k}$) is used to find the value of current air mass flow rate ($\boldsymbol{u}^{k}$) using (\ref{eq18}). Finally, $\boldsymbol{u}^{k}$ is implemented to control the HVAC system, which results in evolution of the system based on (\ref{eq1}), (\ref{eq2}) and (\ref{eq6}). The measured values of the states ($\boldsymbol{x}^{k+1}$) are used as the initial values in the next sampling time and the algorithm continues.

\section{Simulation results}
\label{Simulation results}
In this section, we conduct a set of experiments to validate the accuracy and efficiency of the proposed LMPC controller by bench-marking the results against an equivalent nonlinear controller. For the thermal building model, we consider a thermal zone with 7 states (four states for temperature of walls, two states for temperature of floor and ceiling, and one state for indoor thermal zone temperature) with the parameters the same as \cite{liu2018coordinating} and \cite{haghighi2011modeling}. Other building parameters are:  $d_p=0$, $T_s=10$ ($^{\circ}C$), $P_{rated}$ and $u_{rated}$ are $600 W$ and $1 kg/s$, respectively. The simulations are performed on a dual core i7 3.41 GHz processor with 16 GB of RAM.

\begin{figure}[!t]
     \centering
    \subfloat[Ambient Temperature\label{fig:Tout}]{
		\includegraphics[width=0.45\linewidth]{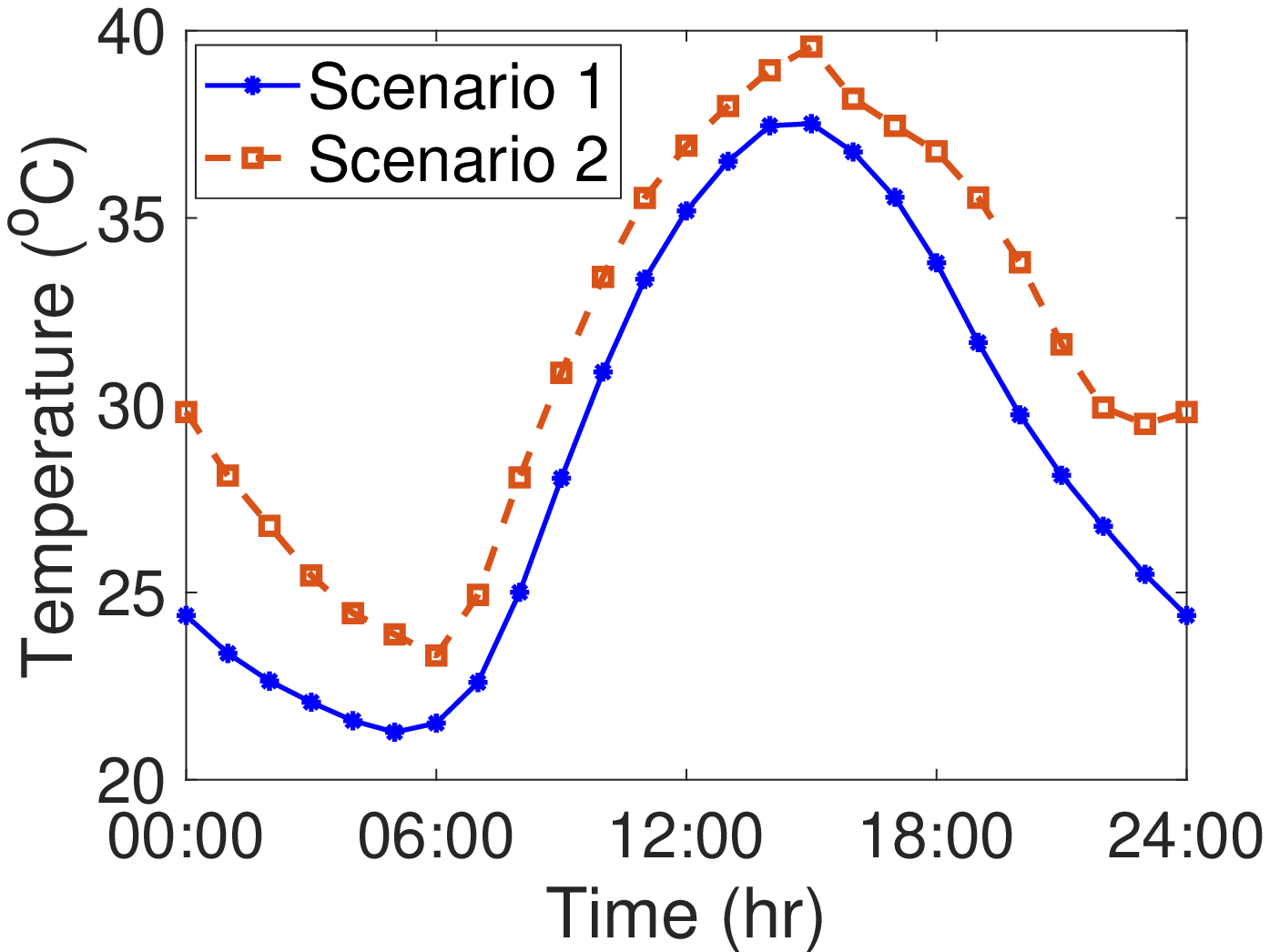}}
		\hfill
	    \subfloat[TOU electricity tariffs\label{fig:TOU}]{
		\includegraphics[width=0.45\linewidth]{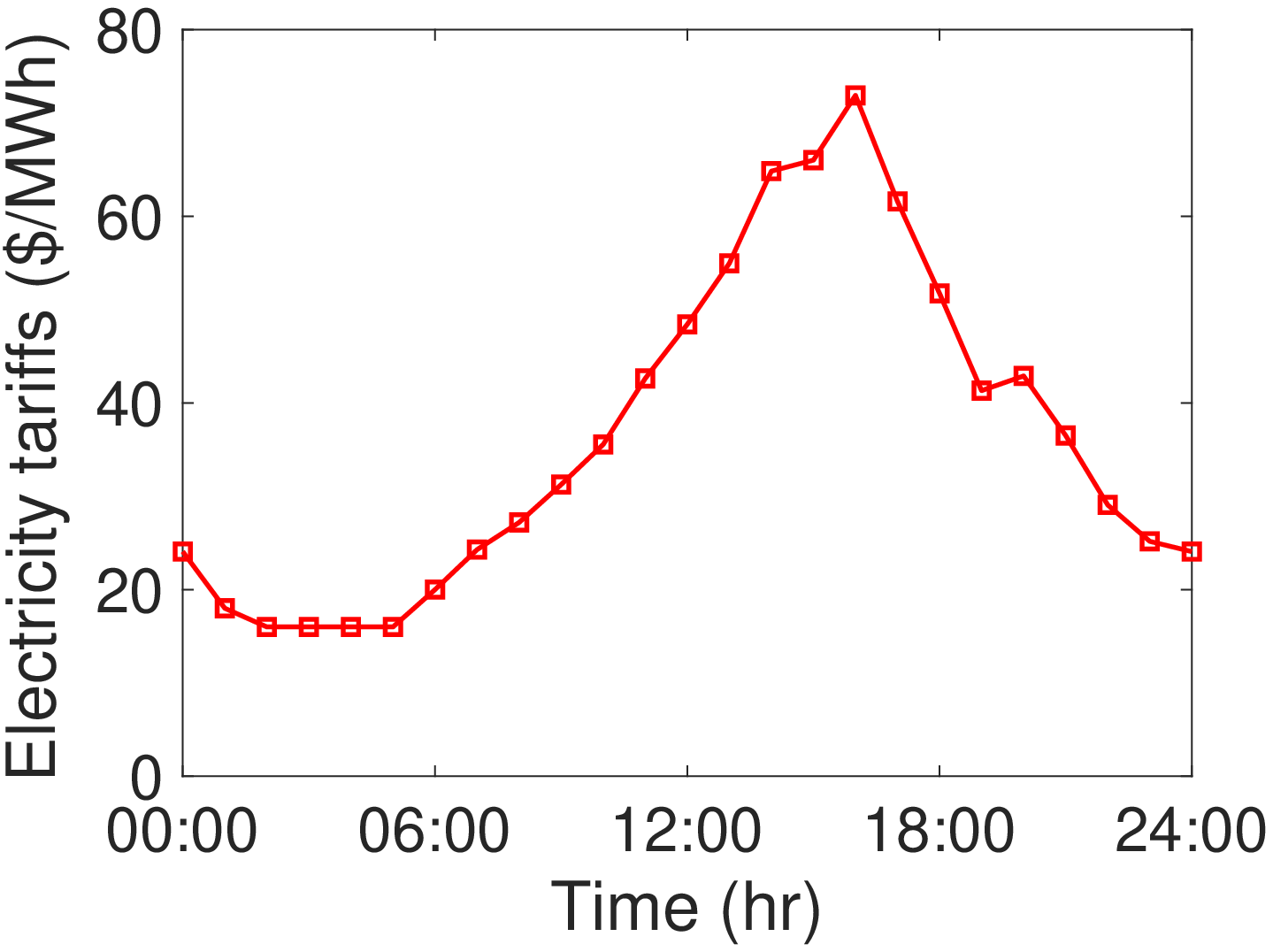}}
		
\subfloat[Received occupancy patterns at the beginning of the day\label{fig:213}]{
\centering
\resizebox{8.5cm}{!}{%
\begin{tabular}{llllllllllllll}
\cline{1-13}
\multicolumn{1}{l|}{Occupancy pattern 1} & \multicolumn{1}{l|}{\cellcolor[HTML]{000000}{\color[HTML]{000000} }} & \multicolumn{1}{l|}{\cellcolor[HTML]{000000}{\color[HTML]{000000} }} & \cellcolor[HTML]{000000}{\color[HTML]{000000} } & \multicolumn{1}{l|}{} & \multicolumn{1}{l|}{} & \multicolumn{1}{l|}{} & \multicolumn{1}{l|}{} & \multicolumn{1}{l|}{} & \multicolumn{1}{l|}{} & \multicolumn{1}{l|}{\cellcolor[HTML]{000000}} & \multicolumn{1}{l|}{\cellcolor[HTML]{000000}} & \multicolumn{1}{l|}{\cellcolor[HTML]{000000}} &    \\ \cline{1-13}
\multicolumn{2}{r}{00:00} &  & \multicolumn{2}{l}{06:00} &  & \multicolumn{2}{l}{12:00} &  & \multicolumn{2}{l}{18:00} & \multicolumn{1}{r}{} & \multicolumn{2}{l}{24:00} \\
 &  &  &  & \multicolumn{3}{l}{Time (hr)} &  &  &  &  &  &  & 
\end{tabular}%
}}
\caption{Model Parameters for Simulations}
\label{fig:3}
\vspace{-0.3 cm}
\end{figure}

The predicted ambient temperatures, the 24-hour TOU electricity tariffs and occupancy patterns for the building received at the beginning of the day are shown in Fig. \ref{fig:3}. The occupancy pattern represents the typical scenarios for a residential building. Simulations are carried out for two different scenarios that correspond to two separate days with different ambient temperatures. To maintain the desired comfort level for buildings' occupants, it is assumed that during occupancy, the indoor temperature in building should lie between 21- 25 ($^{\circ}C$), otherwise, there is no limit for the thermal zone temperatures. There are no temperature limits for the other 6 states of the thermal zones. 
Starting at the beginning of a day (00:00) and after receiving one-day ahead information, the optimization problem is solved using the proposed LMPC for the next 24 hours at a sampling rate of 15 minutes as described in Section \ref{LMPC}. As it is common in MPC-based problems, the length of prediction window $W$ (e.g. $24\times4$ here) is same at every sampling time. For comparison, the original nonlinear optimization problem with nonlinear thermal dynamics and the HVAC system power consumption equations is also solved using an equivalent NLMPC. 
\begin{figure}[!t]
     \centering
    \subfloat[Room temperature-Scenario 1\label{T 1}]{
		\includegraphics[width=0.485\linewidth]{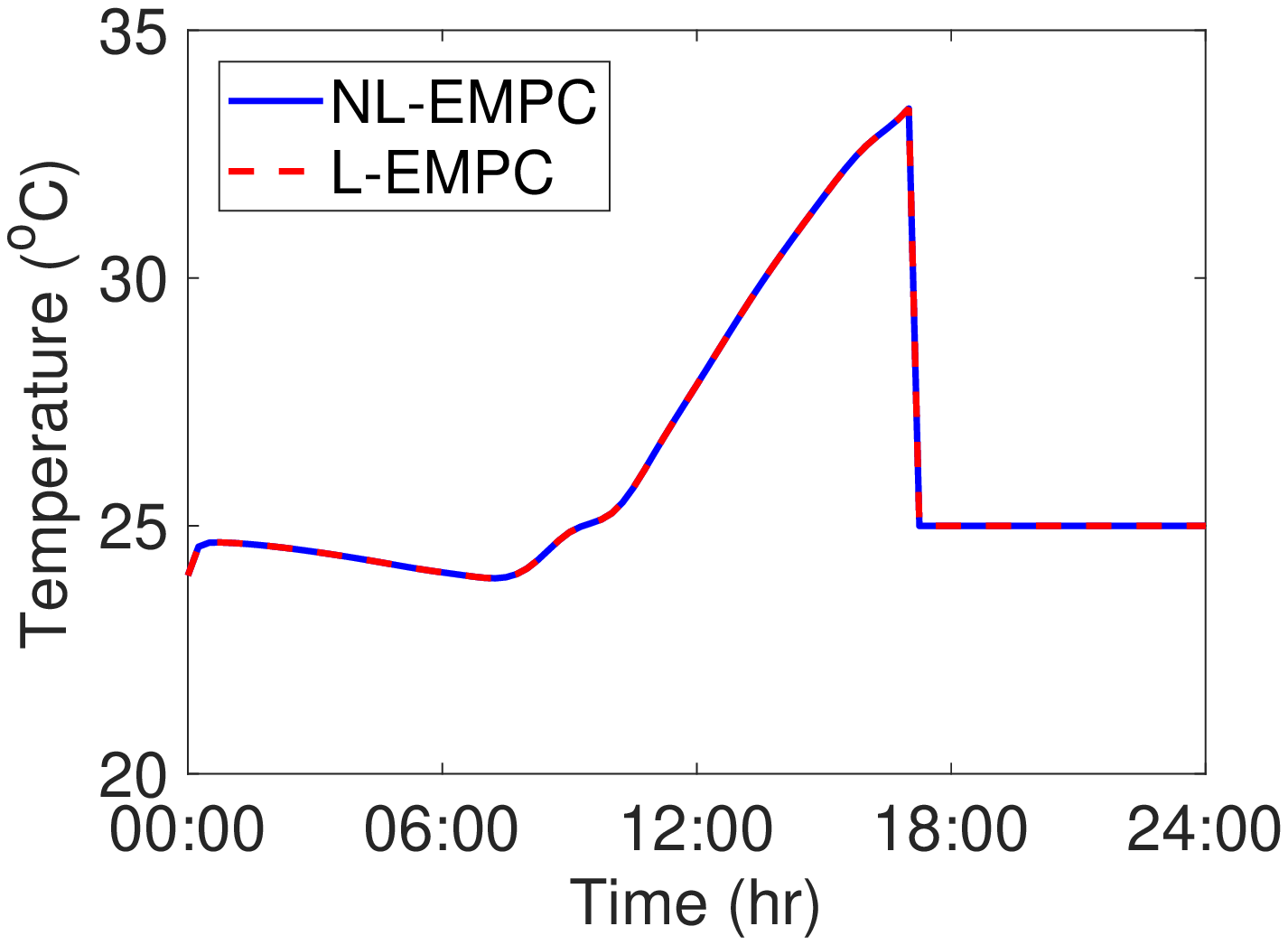}}
	    \hfill
	\subfloat[Room temperature-Scenario 2\label{T 2}]{
		\includegraphics[width=0.485\linewidth]{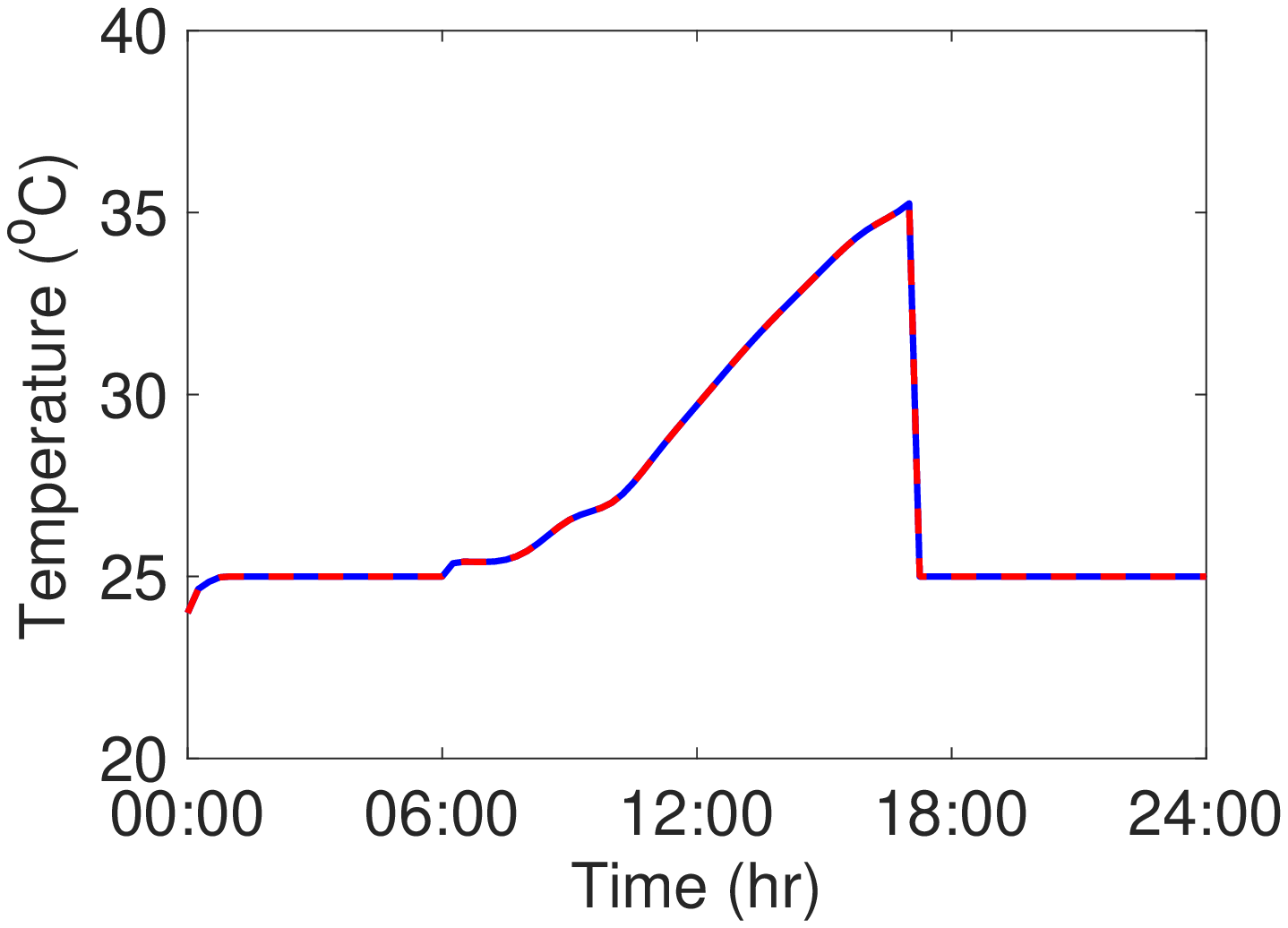}}
		\vspace{-0.2 cm}
	\\   
	    \subfloat[Air mass flow-Scenario 1\label{u 1}]{
		\includegraphics[width=0.485\linewidth]{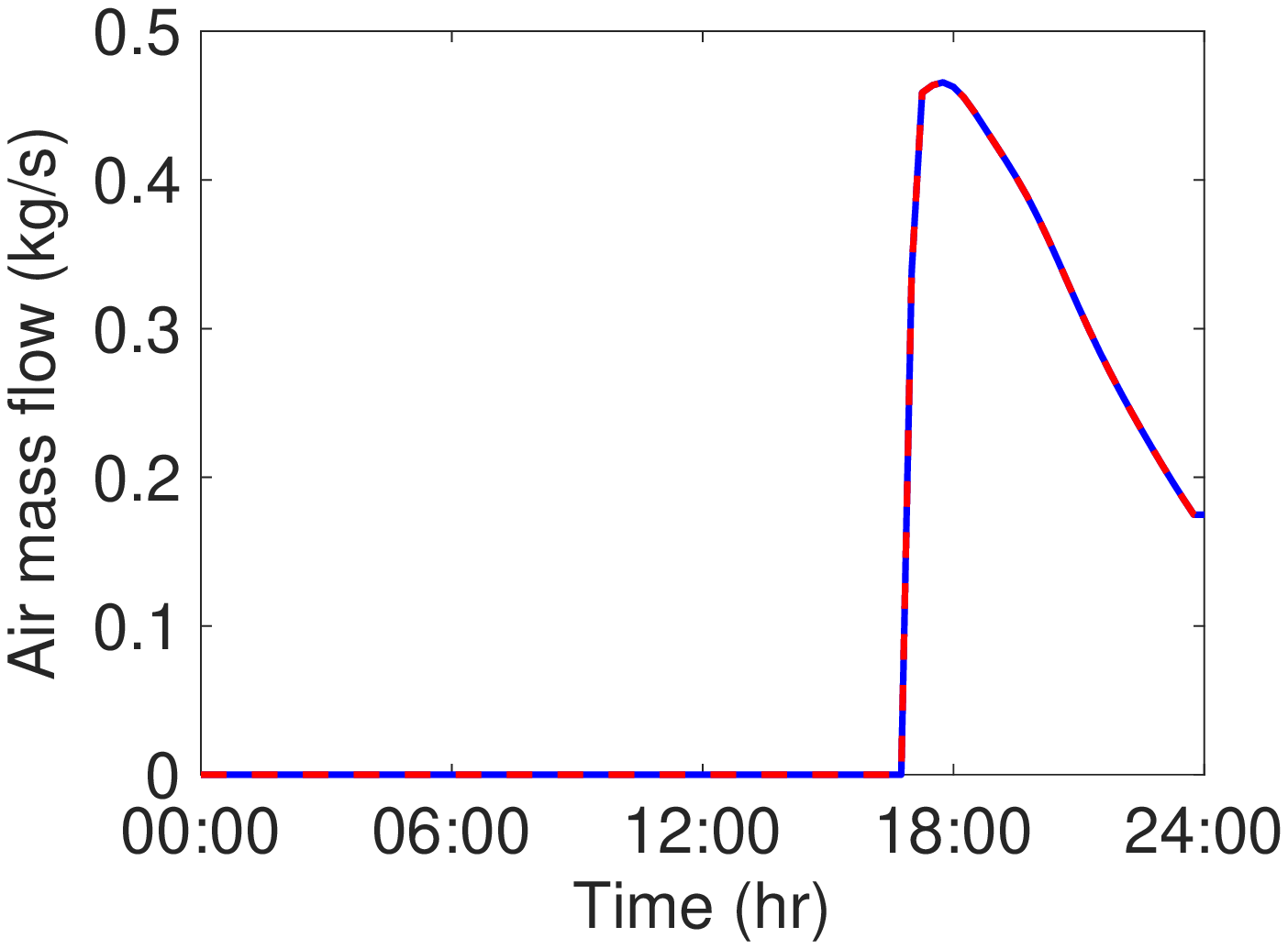}}
	    \hfill
	\subfloat[Air mass flow-Scenario 2\label{u 2}]{
		\includegraphics[width=0.485\linewidth]{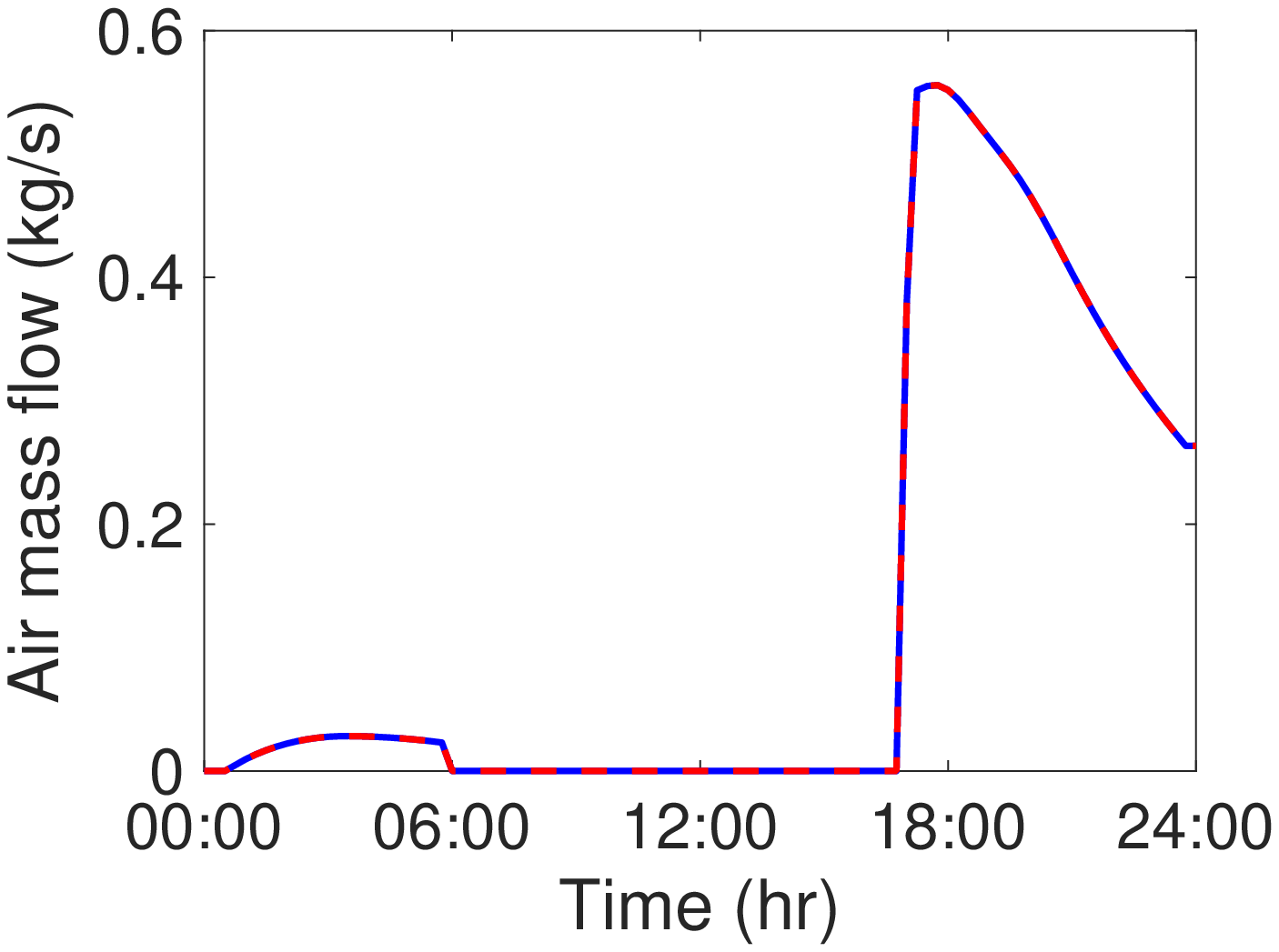}}
		\vspace{-0.2 cm}
			\\   
	    \subfloat[HVAC power-Scenario 1\label{p 1}]{
		\includegraphics[width=0.485\linewidth]{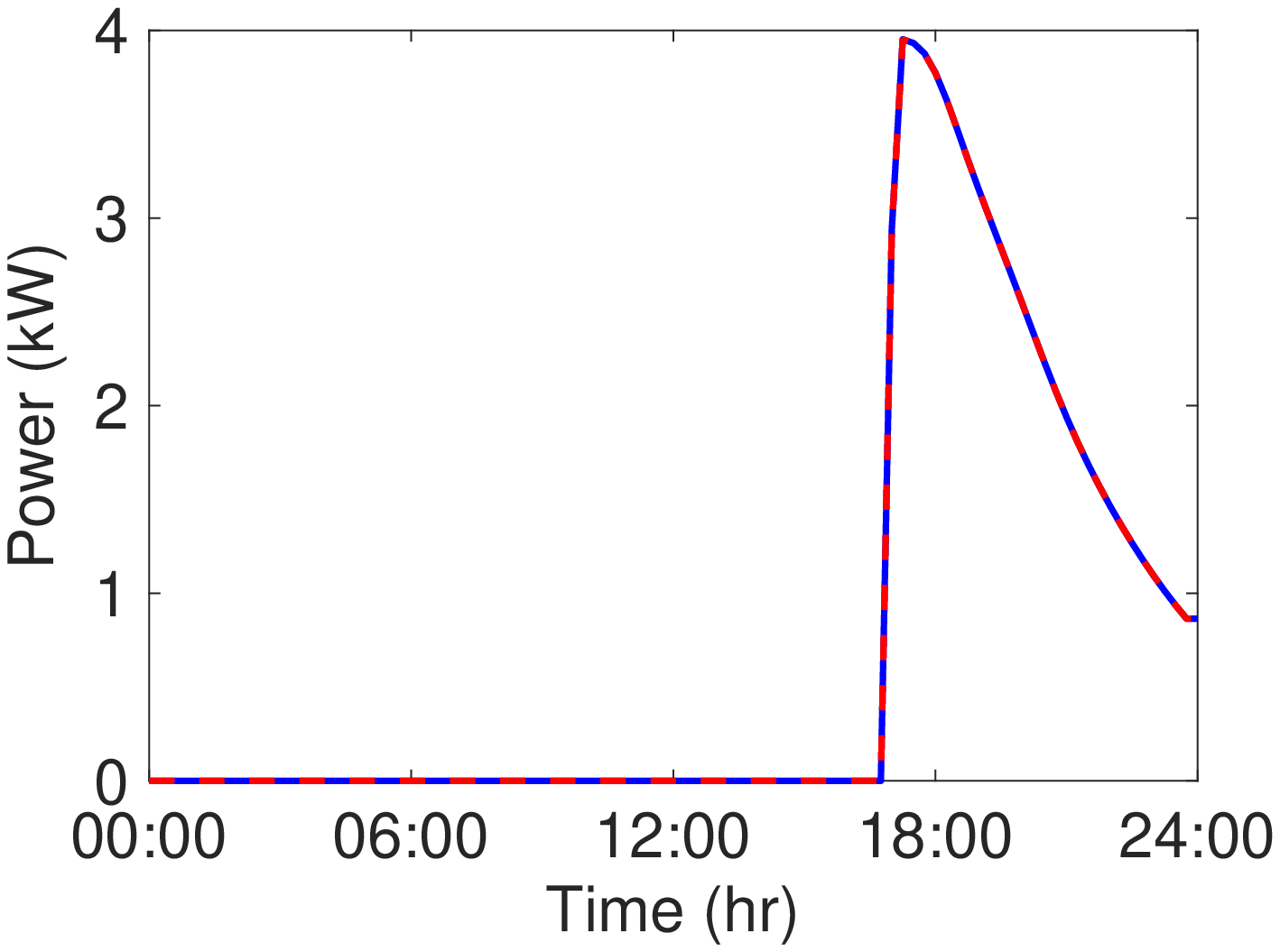}}
	    \hfill
	\subfloat[HVAC power-Scenario 2\label{p 2}]{
		\includegraphics[width=0.485\linewidth]{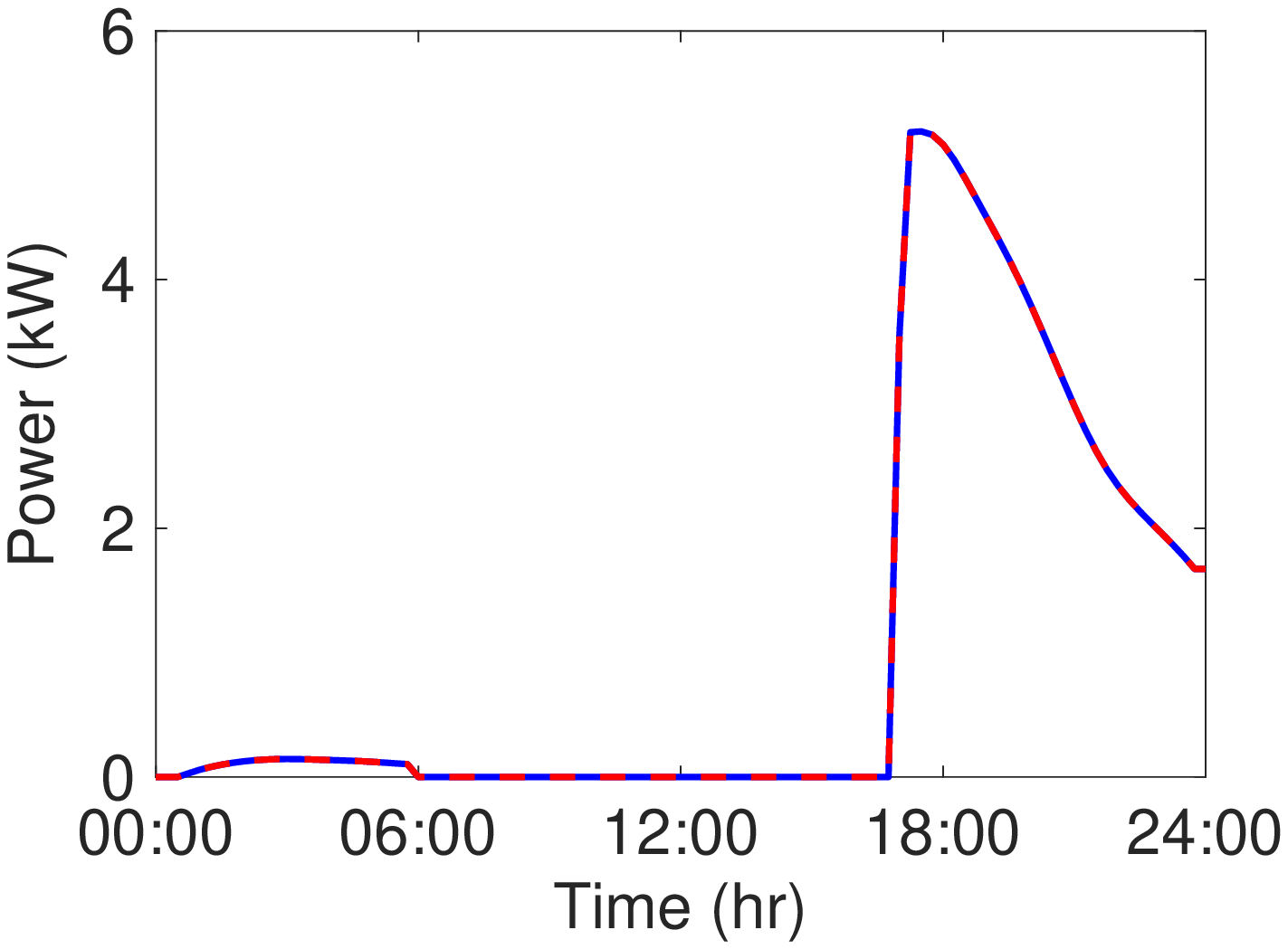}}
  \caption{Comparison of nonlinear and LMPC controllers based on internal zone temperature, air mass flow and HVAC power consumption for defined scenarios.}
\label{fig:5}
\vspace{-0.5cm}
\end{figure}

Fig. \ref{fig:5} shows the evolution of indoor building temperature,  optimal value of the control input $u$ (air mass flow rate), and HVAC power consumption for both scenarios. As is seen in Fig. \ref{u 1} and \ref{u 2}, when the building is occupied, the controllers adjust the air mass flow rate of the HVAC cooling system such that the temperature of the building lies within the prespecified comfort range (see Fig. \ref{T 1}) while simultaneously minimizing the cost of transacted energy. On the contrary, when there is no occupancy in the building, controllers minimize the total cost of energy usage by turning the HVAC cooling system off. Note that there are times during the day (e.g. 00:00-06:00 in scenario 1) that although the building is occupied, there is no need to utilize HVAC ($u=P_H=0$) system. That is, the ambient temperature at these times are low and sufficient to maintain the thermal-zone temperature within the occupants' comfort level without requiring the HVAC cooling system. As it can be seen, LMPC can leverage the thermal building dynamics to minimize the overall cost of transacted energy while approximating the nonlinear dynamics.
\begin{table}[t]
\centering
\caption{Simulation details of LMPC and NLMPC}
\vspace{-0.3cm}
\label{table1}
\begin{tabular}{|l|c|c|c|}
\hline
Approach  & Cost(\$) & Time(s) & Solver \\ \hline
NLMPC  & 0.7286 & 6109 & MATLAB-IPOPT \\ \hline
LMPC  & 0.7286 & 35 & MATLAB-linprog 
  \\ \hline
\end{tabular}%
\vspace{-0.55cm}
\end{table}

Next, Table \ref{table1} presents a comparison of LMPC and an equivalent nonlinear MPC (NLMPC) in terms of optimal cost and computation time for both scenarios. We use IPOPT and linprog functions in MATLAB to solve a nonlinear and LMPC models, respectively. As is seen, there is a negligible difference in cost function obtained using nonlinear and LMPC controllers while the LMPC controller significantly improves the simulation time. Therefore, it is concluded that the LMPC closely mimics the behaviour of the nonlinear plant model. Furthermore, the LMPC leads to approximately 200 times improvement in the computation speed, making it more suitable for real-time control of buildings' HVAC systems.

	
\vspace{-0.3cm}	
\section{Conclusions}
\label{Conclusion and future works}
This paper presents a computationally efficient linear model-predictive controller (LMPC) to optimize the operation of buildings’ HVAC systems. The proposed framework is based on feedback linearization technique when bilinear thermal dynamics are transformed to a linear dynamical model. Next, using prediction-based constraint mapping and piecewise linearization approaches, we transform the nonlinear constraints for the HVAC system power consumption. Note that unlike Jacobian linearization, we do not make any assumption regarding the desired temperature trajectory for the building thus allowing for large variations in indoor temperature especially when building is not occupied. Thus, the proposed linearized model is able to optimize for a system-level objectives that requires leveraging buildings' occupancy information. We detail two such optimization problems viz. minimization of transacted cost of electricity and budget-constrained maximization of occupants’ comfort. Using the proposed linearized approach, we demonstrate that the LMPC performs as well as an equivalent nonlinear MPC (NLMPC) while providing a significant computational advantage. The added computational efficiency makes the proposed LMPC suitable for the real-time optimal control of buildings' HVAC systems. 	
	\bibliographystyle{ieeetr}
	\bibliography{references}

\end{document}